\documentclass[12pt,titlepage,final]{article}

\begin{document}
  \begin{titlepage}
    \begin{flushright}
{\small\sf LPM/02-11\\
      UPRF-2002-11}
    \end{flushright}
\vskip 1.in
    \begin{center}
\textbf{\Large Boundary One-Point Functions, Scattering, and
Background Vacuum Solutions in Toda Theories}\\[2.em] \textbf{\large
V. A. Fateev\footnote{Laboratoire de Physique Math\'{e}matique,
Universit\'{e} Montpellier II, Pl. E.Bataillon, 34095 Montpellier,
France.}$^,$\footnote{On leave of absence from Landau Institute for
Theoretical Physics, ul.Kosygina 2, 117940 Moscow, Russia.}
\hspace{0.2cm} and E. Onofri\footnote{Dipartimento di Fisica,
Universit\`a di
Parma, and {\small\sf I.N.F.N.}, Gruppo Collegato di Parma, 43100 Parma,
Italy}
}\\[5.em]
\end{center}

\date{}

\bigskip
\bigskip

\begin{center}
\textbf{Abstract}
\end{center}
\bigskip
The parametric families of integrable boundary affine Toda theories are
considered. We calculate boundary one-point functions and propose boundary
$S$-matrices in these theories. We use boundary one-point functions and
$S$-matrix amplitudes to derive boundary ground state energies and exact
solutions describing classical vacuum configurations.

 \end{titlepage}

\section{ Introduction}

There is a large class of massive 2D integrable quantum field theories
(QFTs) which can be considered as perturbed conformal field theories (CFTs).
The short distance behavior of these QFTs is encoded in the CFT data while
their long distance properties are determined by $S-$matrix data. A link
between these two kinds of data would provide a good viewpoint for a rather
complete understanding and description of such theories.

In this paper we study the consistency conditions between CFT and
$S$-matrix data for integrable families
of boundary non-simply laced affine Toda
theories (ATTs). The similar analysis for simply laced ATTs (where
integrable boundary conditions do not contain parametric families) was done
in Ref.\cite{FO}. These QFTs can be considered as perturbed CFTs
(non-affine Toda theories),
 which possess an extended symmetry generated by
$W-$algebra. The integrable boundary conditions preserve
this symmetry. This
permits us to apply the ``reflection amplitude'' approach \cite{FLZ} for the
calculation of \ the boundary vacuum expectation values (VEVs) in ATTs. The
boundary VEVs (one-point functions), in particular, contain the information
about the boundary values of the solutions of classical boundary Toda
equations. The information about the long distance behavior of these
solutions can be extracted from boundary $S-$matrix. The explicit solutions
constructed in this paper provide us the consistency check of CFT and
$S$-matrix data in integrable families of boundary ATTs.

The plan of the paper is as follows: in section 2 we recall some basic
facts about Toda theories and one-point functions in ATTs defined on
the whole plane. In section 3 we consider integrable boundary ATTs. We
give the explicit expressions for boundary one point functions which
can be derived by the ``reflection amplitude'' approach
\cite{FLZ}. These functions determine, in particular, the boundary
values of the solutions corresponding to the vacuum configurations in
ATTs. We use the boundary VEVs to derive a conjecture for quantum
boundary ground state energies. In section 4 we construct the boundary
scattering theory which is consistent with this conjecture. The
semiclassical limits of boundary $S-$matrix amplitudes fix the
asymptotics of vacuum solutions in ATTs. These asymptotics determine
completely the explicit form of the solutions which we construct in
section 5.  The solutions can be written in terms of tau-functions
associated with multisoliton solutions in ATTs. We check that the
boundary values of these solutions agree with the corresponding
values given by  boundary one-point functions.

\section{ Affine and non-affine Toda theories}

The ATT corresponding to a Lie algebra $G$ of rank $r$ is described by the
action\footnote{In this paper we denote as $G$ also the corresponding
untwisted
affine algebra and use the notation $G^{\vee}$ for dual affine algebra.}

\begin{equation}
\mathcal{A}_{b}=\int d^{2}x\left[ \frac{1}{8\pi }(\partial _{\mu }\varphi
)^{2}+\sum_{i=0}^{r}\mu _{i}e^{be_{i}\cdot \varphi }\right] ,  \label{at}
\end{equation}
where $e_{i},$ $i=1,...,r$ are the simple roots of Lie algebra $G$
and $-e_{0}$ is a maximal root which defines the integers $n_{i}$
by the relation:

\begin{equation}
\sum_{i=0}^{r}n_{i}e_{i}=0.  \label{mr}
\end{equation}
The fields $\varphi $ in eq. (\ref{at}) are normalized so that at $\mu
_{i}=0 $

\begin{equation}
\left\langle \varphi _{a}(x) \varphi _{b}(y)\right\rangle =-\delta _{ab}\log
\left| x-y\right| ^{2}  \label{n}
\end{equation}

We consider below mainly the non-simply laced ATTs. These theories possess
a remarkable property of duality \cite{DGZ}.
Under duality transformation $b\rightarrow b^{\vee }=1/b$
the action (\ref{at}) transforms to the action
of dual ATT characterized by the roots $e_{i}(G^{\vee
})=2e_{i}(G)/e_{i}^{2}(G)$. The parameters $\mu _{i}$ in the dual ATTs are
related as follows:

\begin{equation}
\pi \mu _{i}(G)\gamma (e_{i}^{2}b^{2}/2)=\left( \pi \mu _{i}(G^{\vee
})\gamma (2/(e_{i}^{2}b^{2}))\right) ^{e_{i}^{2}b^{2}/2}  \label{mumu}
\end{equation}
where as usual $\gamma (z)=\Gamma (z)/\Gamma (1-z)$.

The mass ratios in non-simply laced ATT are different from classical ones
\cite{DGZ}. For real $b$ the spectrum consists of $r$ particles with masses
$m_{j}$. These masses of particles
are characterized by one mass parameter $m$
and flow from the values $m\lambda _{j}(G)$ to the values $m\lambda
_{j}(G^{\vee })$, where $\lambda _{j}^{2}(G)$ and $\lambda _{j}^{2}(G^{\vee
})$ are the eigenvalues of \ classical mass matrices $M(G)$ and $M(G^{\vee
}) $, i.e.:

\begin{equation}
M_{ab}=\sum_{i=0}^{r}n_{i}(e_{i})^{a}(e_{i})^{b}.  \label{mab}
\end{equation}
To describe the spectrum it is convenient to introduce the notations:

\begin{equation}
x=\frac{b^{2}}{1+b^{2}};\qquad h=H(G)(1-x)+H(G^{\vee })x  \label{ah}
\end{equation}
where $H$ is the Coxeter number. Then the spectrum can be expressed in terms
of parameter $m$ as:

\begin{eqnarray}
B_{r} &:&\quad m_{r}=\sqrt{2}m,~m_{j}=\sqrt{8}m\sin (\pi j/h);~j=1,...,r-1.
\nonumber \\
C_{r},BC_{r} &:&\quad m_{j}=\sqrt{8}m\sin (\pi j/h);~j=1,...,r  \label{spr}
\end{eqnarray}
The spectrum for the exceptional algebras $G$ can be found in \cite{DGZ}.

The exact relation between the parameter $m$ in the above spectra and the
parameters $\mu _{i}$ in the action (\ref{at}) can be obtained by the Bethe
ansatz method (see for example Refs. \cite{ALZ},\cite{F}). It was derived in
Ref.\cite{ABF} and has the form:

\begin{equation}
\prod_{i=0}^{r}\left[ \frac{-\pi \mu _{i}}{\gamma
(-e_{i}^{2}b^{2}/2)}\right]
^{n_{i}}=\left[ \frac{mk(G)\Gamma \left( \frac{1-x}{h}\right) \Gamma \left(
\frac{x}{h}\right) x}{\sqrt{2}\Gamma \left( \frac{1}{h}\right) h}\right]
^{2h(1+b^{2})},  \label{mmu}
\end{equation}
where integers $n_{i}$ are defined by the equation (\ref{mr}) and

\begin{equation}
k(B_{r})=2^{-2/h}\Gamma ,\quad k(C_{r})=2^{2x/h},\quad
k(BC_{r})=2^{(x-1)/h}.
\label{k}
\end{equation}
For the exceptional algebras $G$ the numbers $k(G)$ can be found in \cite
{ABF}. The similar relations for the dual ATTs can be easily derived from
Eqs.(\ref{mmu},\ref{k}) and duality relations (\ref{mumu}).

The ATTs can be considered as perturbed CFTs. With the parameter
$\mu_{0}=0$ the action (\ref{at}) describes non-affine Toda theories
(NATTs), which are conformal field theories. To describe the generator
of conformal symmetry we introduce the complex coordinates
$z=x_{1}+ix_{2},\ \bar{z}=x_{1}-ix_{2}$ and vector

\begin{equation}
Q=\rho /b+b\rho ^{\vee };\qquad \rho =\frac{1}{2}\sum_{\alpha >0}\alpha
,\quad \rho ^{\vee }=\sum_{\alpha >0}\alpha /\alpha^{2}. 
\label{q}
\end{equation}
where the sum in the definition of the Weyl vector $\rho $ runs over all
positive roots $\alpha $ of $G.$

The holomorphic stress energy tensor

\begin{equation}
T(z)=-\frac{1}{2}(\partial _{z}\varphi )^{2}+Q\cdot \partial _{z}^{2}\varphi
\label{se}
\end{equation}
ensures the local conformal invariance of the NATT. Besides conformal
invariance NATT possesses an additional symmetry generated by two copies of
the chiral $W(G)$-algebras: $W(G)\otimes $ $\overline{W}(G)$ (see Refs.
\cite
{FL},\cite{VFT} for details). This extended conformal symmetry permits to
calculate the VEVs of exponential fields in ATTs, i.e. one point functions

\begin{equation}
\mathbf{G}(a)=\left\langle 0|\exp (a\cdot \varphi )|0\right\rangle .
\label{G}
\end{equation}
These VEVs were calculated in \cite{ABF}. They can be represented in the
form:

\begin{equation}
\mathbf{G}(a)=N_{G}^{2}(a)\exp \left( \int\limits_{0}^{\infty }
\frac{dt}{t}[a^{2}e^{-2t}-\sum_{\alpha >0}\sinh ((\alpha ^{2}b^{2}/2+1)t)
F_{\alpha }(a,t)]\right) .  \label{GG}
\end{equation}
In this equation the factor $N_{G}(a)$ can be expressed in terms of
fundamental co-weights $\omega _{i}^{\vee }$ of $G$
(i.e. $\omega _{i}^{\vee}\cdot e_{j}=\delta _{ij}$) as:

\begin{equation}
N_{G}^{2}(a)=\left[ \frac{mk(G)\Gamma \left( \frac{1-x}{h}\right) \Gamma
\left( \frac{x}{h}\right) x}{\sqrt{2}\Gamma \left( \frac{1}{h}\right) h}
\right] ^{2Q\cdot a-a^{2}}\prod_{i=1}^{r}\left[ \frac{-\pi \mu _{i}}{\gamma
\left( -e_{i}^{2}b^{2}/2\right) }\right] ^{-\omega _{i}^{\vee }\cdot a}
\label{NA}
\end{equation}
and the function $F_{\alpha }(a,t)$ is given by

\begin{equation}
F_{\alpha }(a,t)=\frac{\sinh (ba_{\alpha }t)\,\sinh (ba_{\alpha 
}-2bQ_{\alpha
}+h(1+b^{2}))t)}{\sinh t\,\sinh (\alpha ^{2}b^{2}t/2)\,\sinh ((1+b^{2})ht)}
\label{Fa}
\end{equation}
(here and below the subscript $\alpha $ denotes the scalar product of the
vector with a positive root $\alpha $, i.e. $a_{\alpha }=a\cdot \alpha $;
$Q_{\alpha }=Q\cdot \alpha $ and so on).

This expression satisfies many possible perturbative and non-perturbative
tests for one-point function in ATT. For example, it can be easily derived
from Eqs.(\ref{at},\ref{mmu}) that the bulk vacuum energy $E(G)$ can be
expressed in terms of function $\mathbf{G}(a)$. We have:

\begin{equation}
\mu _{i}\partial _{\mu _{i}}E=n_{i}(1-x)E/h=\mu _{i}\mathbf{G}(be_{i}).
\label{EB}
\end{equation}
The values of the function $\mathbf{G(}a\mathbf{)}$ at the special points
$be_{i}$ can be calculated explicitly and the result coincides with the
expression which was obtained by Bethe ansatz method \cite{ABF}. In
particular, for $B,C$ and $BC$ ATT we obtain that

\begin{equation}
E=\frac{m^{2}\sin (\pi /h)}{4\sin (\pi x/h)\sin (\pi (1-x)/h)}. 
\label{BCBC}
\end{equation}

It is convenient to define the field $\widehat{\varphi }=\varphi
b/(1+b^{2})$.
In the limit $b\rightarrow 0$ (as well as in the dual limit $b^{\vee
}=1/b\rightarrow 0$) this field can be described by the classical equations.
In particular, it can be derived from Eqs.(\ref{GG},\ref{mmu}) that the VEV
of this field $\widehat{\varphi }_{0}=\left\langle 0|b\varphi
/(1+b^{2})|0\right\rangle $ in that limit can be expressed in terms of
fundamental co-weights in the form:

\begin{equation}
\widehat{\varphi }_{0}=\sum_{i=1}^{r}\omega _{i}^{\vee }
\log \left( \frac{m^{2}}{4\pi b^{2}\mu _{i}}\right)  \label{vcl}
\end{equation}
and it coincides with a classical vacuum of ATT (\ref{at}). After
rescaling and
shifting of the field $\varphi$, the action (\ref{at}) can be written in
terms of the field $\phi =\widehat{\varphi }-\widehat{\varphi }_{0}$ as the
classical action of ATT:

\begin{equation}
\mathcal{A}_{b}=\frac{1}{4\pi b^{2}}\int d^{2}x[\frac{1}{2}(\partial _{\mu
}\phi )^{2}+m^{2}\sum_{i=0}^{r}n_{i}e^{e_{i}\cdot \phi }]+O(1)  \label{ACL}
\end{equation}
In a similar way, the opposite (dual) limit $b^{\vee }=1/b\rightarrow
0$ leads to the classical action of the dual affine Toda theory.

\section{Boundary Toda Theories, Boundary One-Point Functions and Classical
Vacuum Solutions}

In the previous section we considered Toda theories defined on the whole
plane $R^{2}$. Here we consider these theories defined
on the half-plane $H=(x,y;y>0)$ with integrable boundary conditions.
The action of the boundary
ATTs can be written in the form:

\begin{equation}
\mathcal{A}_{bound}=\int\limits_{H}d^{2}x\left[ \frac{1}{8\pi }(\partial
_{\mu }\varphi )^{2}+\sum_{i=0}^{r}\mu _{i}e^{be_{i}\cdot \varphi }\right]
+\int dx\sum\limits_{0}^{r}\nu _{i}e^{be_{i}\cdot \varphi /2}.  \label{atb}
\end{equation}
These theories also possess the property of duality. Under the duality
transformation $b\rightarrow b^{\vee }=1/b$ they transform to the QFTs which
can be described by the action of the dual ATTs with the parameters $\mu
_{i}(G^{\vee })$ given by Eq.(\ref{mumu}) and boundary parameters $\nu
_{i}(G^{\vee })$ which can be determined in the following way \cite{FZZA}:
we define the variables $s_{0},...,s_{r}$ by the relation:

\begin{equation}
\nu _{i}^{2}(G)=\mu _{i}(G)\frac{\cos ^{2}(\pi l_{i}s_{i}b/2)}{\sin (\pi
l_{i}^{2}b^{2}/2)};\qquad l_{i}=\sqrt{e_{i}^{2}(G)}  \label{nB}
\end{equation}
then the dual boundary parameters are determined as follows:

\begin{equation}
\nu _{i}^{2}(G^{\vee })=\mu _{i}(G^{\vee })\frac{\cos ^{2}(\pi
l_{i}^{\vee } s_{i}b^{\vee }/2)}{\sin (\pi (l_{i}^{\vee }b^{\vee
})^{2}/2)};\qquad l_{i}^{\vee }=\sqrt{e_{i}^{2}(G^{\vee })}=2/l_{i}.
\label{nBD}
\end{equation}
It is easy to derive from these equations that the theory dual to the 
boundary
ATT with the Neumann boundary conditions i.e. all $\nu _{i}=0$ or

\begin{equation}
\partial _{y}\varphi (x,y)|_{y=0}=0  \label{N}
\end{equation}
is specified by the boundary parameters

\begin{equation}
\nu _{i}=\left( \frac{\mu _{i}}{2}\cot (\pi l_{i}^{2}b^{2}/4)\right) ^{1/2}.
\label{ND}
\end{equation}

The integrability conditions for classical boundary ATTs were studied in
Ref. \cite{BCD}. In particular it was shown there that contrary to simply
laced case, the non-simply laced $B_{r},B_{r}^{\vee },C_{r},C_{r}^{\vee }$
and $BC_{r}$ ATTs\footnote{$B_{r}^{\vee }$, $C_{r}^{\vee }$ and $BC_{r}$
ATTs
are known also as $A^{(2)}_{2r-1}$, $D^{(2)}_{r+1}$
and $A^{(2)}_{2r}$ Toda theories}
admit  parametric families of integrable boundary
conditions. These parameters are associated with $\nu _{i}$ corresponding to
the non-standard roots ($e_{i}^{2}\neq 2$) of affine Lie algebras. It is
convenient to write $\nu _{i}^{(cl)}$ in the form:

\begin{equation}
\nu _{i}^{(cl)}=\varepsilon _{i}\sqrt{\frac{\mu _{i}}{\pi b^{2}}}A_{i}
\label{sig}
\end{equation}
where $\varepsilon _{i}=\pm 1$. Then integrable boundary conditions can be
specified by continuous parameters $w_{0}$ and $w_{r}$ as follows:

\begin{eqnarray}
B_{r} &:&e_{r}^{2}=1;\quad A_{i}=1,i\neq r,\quad A_{r}=
\sqrt{2}\cos (\pi w_{r}/2).  \label{Bcl} \\
B_{r}^{\vee } &:&e_{r}^{2}=4;\quad A_{i}=0,i\neq r,\quad A_{r}=
\cos (\pi w_{r})/\sqrt{2}.  \label{Bdcl} \\
C_{r} &:&e_{0,r}^{2}=4;\quad A_{i}=0,i\neq 0,r,\quad A_{0,r}=
\cos (\pi w_{0,r})/\sqrt{2}.  \label{Ccl} \\
C_{r}^{\vee } &:&e_{0,r}^{2}=1;\quad A_{i}=1,i\neq 0,r,\quad A_{0,r}=
\sqrt{2}\cos (\pi w_{0,r}/2).  \label{Cd}
\end{eqnarray}
In the case of $BC_{r}$ ATT ($e_{0}^{2}=1,e_{r}^{2}=4$) there exist two
types of integrable boundary conditions:

\begin{equation}
A_{0}=\sqrt{2}\cos (\pi w_{0}/2),\quad A_{i}=1,i\neq 0,r,\quad A_{r}=
1/\sqrt{2}\quad  \label{BC1}
\end{equation}
and

\begin{equation}
A_{i}=0,i\neq r,\quad A_{r}=\cos (\pi w_{r})/\sqrt{2}.  \label{BC2}
\end{equation}
In this paper we consider the case when in Eq.(\ref{sig}) all signs
$\varepsilon _{i}=1$. The cases corresponding to different signs of
parameters $\varepsilon _{i}$ are more subtle and will be considered in the
separate publication.

The quantum version of these integrable boundary conditions can be described
in the following way. For $B_{r}$($B_{r}^{\vee }$) ATT the boundary
parameters $\nu _{i}(B_{r}),i\neq r$ are defined by Eq.(\ref{ND})
($\nu_{i}(B_{r}^{\vee })=0,i\neq r$
for the $B_{r}^{\vee }$ case) and for $i=r$:

\begin{eqnarray}
\nu _{r}(B_{r}) &=&\cos (\pi s_{r}b/2)\sqrt{\mu _{r}(B_{r})/\sin (\pi
b^{2}/2)};  \nonumber \\
\nu _{r}(B_{r}^{\vee }) &=&\cos (\pi s_{r}b^{\vee })\sqrt{\mu
_{r}(B_{r}^{\vee })/\sin (2\pi (b^{\vee })^{2})}.  \label{BQT}
\end{eqnarray}
For $C_{r}$($C_{r}^{\vee }$) ATT the boundary parameters $\nu
_{i}(C_{r})=0,i\neq 0,r$ ($\nu _{i}(C_{r}^{\vee })$ for $i\neq 0,r$ are
defined by Eq.(\ref{ND}) with the substitution $b\rightarrow b^{\vee }$) and
for $i=0,r$:

\begin{eqnarray}
\nu _{0,r}(C_{r}) &=&\cos (\pi s_{0,r}b)\sqrt{\mu _{r}(C_{r})/\sin (2\pi
b^{2})};  \nonumber \\
\nu _{0,r}(C_{r}^{\vee }) &=&\cos (\pi s_{0,r}b^{\vee }/2)\sqrt{\mu
_{r}(C_{r}^{\vee })/\sin (\pi (b^{\vee })^{2}/2)}  \label{CQT}
\end{eqnarray}

The parameters $w_{0}$ and $w_{r}$ in Eqs.(\ref{Bcl}-\ref{Cd}) coincide with
the classical limits of corresponding parameters $s_{i}b$ or
$s_{i}b^{\vee }$.
It is convenient to define the variables $w_{0}$ and $w_{r}$ by the
relation:

\begin{equation}
w_{0}=\frac{s_{0}b}{1+b^{2}},\quad w_{r}=\frac{s_{r}b}{1+b^{2}}  \label{w}
\end{equation}
These variables are useful for the following quantum calculations. Their
dual classical limits coincide with the corresponding limits of parameters
$s_{i}b$ and $s_{i}b^{\vee }$. In the classical equations below we
consider the parameters $w_{i}$ as these limits.

We note that quite different classical integrable
boundary $G$ and $G^{\vee}$ ATTs in the quantum case are
described by the same theory and are
related by a duality transformation ($b\rightarrow 1/b$). The same is true
for $BC_{r}$ boundary ATT where quantum boundary conditions (\ref{nB}) for
$\nu _{0}$ and (\ref{ND}) for $\nu _{i}$ with $i\neq 0$ flow from boundary
conditions (\ref{BC1}) at $b\rightarrow 0$ to boundary conditions
(\ref{BC2}) in the dual limit.

For the analysis of integrable boundary Toda theories it is useful to
consider the vacuum expectation values of the boundary exponential fields.

\begin{equation}
\mathcal{G}(a)=\left\langle \exp (a\cdot \varphi (x,0)/2)\right\rangle _{B}.
\label{GB}
\end{equation}
These one-point functions can be calculated by the reflection amplitudes
method \cite{FLZ} which uses the extended conformal symmetry of the
background boundary CFTs (NATTs). The similar calculations for simply laced
ATTs were described in details in Refs.\cite{FZZA},\cite{VAF},\cite{VFT}.
Here we give the
explicit expressions for the one-point functions (Eq.\ref{GB})
corresponding to dual pairs
$B_{r}$, $B_{r}^{\vee }$and $C_{r}$, $C_{r}^{\vee }$ of integrable boundary
ATTs. The boundary one-point functions in $BC_{r}$ ATT are more involved and
will be published elsewhere.

The one-point function for the dual pair of $\ B_{r}$ and $B_{r}^{\vee
}$ ATTs with integrable boundary conditions described above depends on
one boundary parameter $s_{r}$ and can be represented in the form:

\begin{equation}
\mathcal{G}_{B}(a)=\mathcal{D}_{B}(a)\exp \left( \int\limits_{0}^{\infty }
\frac{dt}{t}(\sinh ^{2}(s_{r}bt)-\sinh ^{2}(b^{2}t/2))f(a,t)\right) .
\label{DB}
\end{equation}
In this equation $\mathcal{D}_{G}(a)$ denotes boundary one-point function in
ATT corresponding to the Lie algebra $G$ \ and characterized by dual
Neumann boundary conditions (\ref{ND}). This function can be written as
follows:

\begin{equation}
\mathcal{D}_{G}(a)=N_{G}(a)\exp \left( \int\limits_{0}^{\infty }\frac{dt}{t}
\left[ \frac{a^{2}}{2}e^{-2t}-\sum_{\alpha >0}\Psi _{\alpha }(t)
\mathcal{F}_{\alpha }(a,t)\right] \right)  \label{DG}
\end{equation}
where the factor $N_{G}(a)$ is defined by Eq.(\ref{NA}) and functions $\Psi
_{\alpha }$ and $\mathcal{F}_{\alpha }$ are:

\begin{equation}
\Psi _{\alpha }(t)=2e^{t}\sinh ((1+\alpha ^{2}b^{2}/2)t)\cosh (\alpha
^{2}b^{2}t/2);  \label{psi}
\end{equation}

\begin{equation}
\mathcal{F}_{\alpha }(a,t)=\frac{\sinh (ba_{\alpha }t)\sinh (ba_{\alpha
}-2bQ_{\alpha }+h(1+b^{2}))t)}{\sinh 2t\sinh (\alpha ^{2}b^{2}t)\sinh
((1+b^{2})ht)}.  \label{cal}
\end{equation}
To define the function $f(a,t)$ in Eq.(\ref{DB}), we denote as
$\mathbf{B}$ the set of positive short roots of the Lie algebra
$B_{r}$ ($\mathbf{B:}\alpha>0,\alpha ^{2}=1$); then this function has
the form:

\begin{equation}
f(a,t)=2\sum_{\alpha \in \mathbf{B}}\mathcal{F}_{\alpha }(a,t).  \label{f}
\end{equation}
To obtain the one point function in the dual integrable boundary
$B_{r}^{\vee }$ ATT we should make the substitution $b\rightarrow 1/b$
in the equations written above and use the duality relations
(\ref{mumu}) for parameters $\mu _{i}$.

The one point function for the dual pair of $\ C_{r}$ and $C_{r}^{\vee
}$ ATTs with integrable boundary conditions (\ref{CQT}) depends on two
boundary parameters $s_{0}$, $s_{r}$ and can be represented in the
form:

\begin{eqnarray}
\mathcal{G}_{C}(a) &=&\mathcal{D}_{C}^{\vee }(a)\exp \left(
\int\limits_{0}^{\infty }\frac{dt}{t}(\sinh ^{2}(2s_{0}bt)-\sinh
^{2}t)f_{0}(a,t)\right)  \nonumber \\
&&\times \exp \left( \int\limits_{0}^{\infty }\frac{dt}{t}(\sinh
^{2}(2s_{r}bt)-\sinh ^{2}t)f_{r}(a,t)\right) .  \label{DC}
\end{eqnarray}

In this equation $\mathcal{D}_{G}^{\vee }(a)$ denotes the one-point function
in ATT corresponding to the Lie algebra $G$ and characterized by Neumann
boundary conditions (\ref{N}). This function is defined by Eq.(\ref{DG})
where we should do the substitution $\Psi _{\alpha }(t)\rightarrow \Psi
_{\alpha }^{\vee }(t)$ with:

\begin{equation}
\Psi _{\alpha }^{\vee }(t)=2\exp (\alpha ^{2}b^{2}t/2)\sinh ((1+\alpha
^{2}b^{2}/2)t)\cosh t.  \label{pd}
\end{equation}
To define the functions $f_{0,r}(a,t)$ in Eq.(\ref{DC}) we denote by
$\mathbf{C}$ the set of positive long roots of Lie algebra $C_{r}$
($\mathbf{C:}\alpha >0,\alpha ^{2}=4$).
Then these functions have a form:

\begin{equation}
f_{0}(a,t)=2\sum_{\alpha \in \mathbf{C}}\frac{\sinh (ba_{\alpha }t)\,\sinh
((ba_{\alpha }-2bQ_{\alpha })t)}{\sinh 2t\sinh (\alpha ^{2}b^{2}t)\,\sinh
((1+b^{2})2ht)},  \label{f1}
\end{equation}

\begin{equation}
f_{r}(a,t)=2\sum_{\alpha \in \mathbf{C}}\frac{\sinh (ba_{\alpha }t)\,\sinh
((ba_{\alpha }-2bQ_{\alpha }+2h(1+b^{2}))t)}{\sinh 2t\,\sinh (\alpha
^{2}b^{2}t)\,\sinh ((1+b^{2})2ht)}.  \label{f2}
\end{equation}
To obtain the one point function in the dual integrable boundary
$C_{r}^{\vee }$ ATT we should do the
substitution $b\rightarrow 1/b$ and to
use the duality relations (\ref{mumu}) for parameters $\mu _{i}$.

To study the classical limit of boundary Toda equations it is convenient to
introduce the vector $\Theta _{b}$ which is equal to the difference of the
boundary VEV of the field $\widehat{\varphi }=b\varphi /(1+b^{2})$ and the
VEV of this field in ATT defined on the whole plane (see section 2):

\begin{equation}
\Theta _{b}(G)=\left\langle \widehat{\varphi }(x,0)\right\rangle
_{B}-\left\langle 0|\widehat{\varphi }|0\right\rangle =\frac{b}{(1+b^{2})}
\partial _{a}(2\mathcal{G}(a)\mathcal{-}\mathbf{G(}a\mathbf{))|}_{a=0}
\label{th}
\end{equation}
where $\mathcal{G}$ denotes $\mathcal{G}_{B}$ or $\mathcal{G}_{C}$ and
$\mathbf{G}$ is defined by Eq.(\ref{GG}). In particular the explicit
expression for $\Theta _{b}(B)$ can be written as follows:

\begin{eqnarray}
\frac{\Theta _{b}(B)}{x(1-x)} &=&-\sum_{\alpha >0}\alpha
\int\limits_{0}^{\infty }\frac{dt\sinh ((1-x+\alpha ^{2}x/2)t)\chi _{\alpha
}(h,t)}{\sinh (\alpha ^{2}xt/2)\,\cosh ((1-x)t)\sinh ht}  \nonumber \\
&&+\sum_{\alpha \in \mathbf{B}}4\alpha \int\limits_{0}^{\infty }
\frac{dt(\sinh ^{2}(w_{r}t)-\sinh ^{2}(xt/2))\chi _{\alpha }(h,t)}
{\sinh (\alpha^{2}xt)\sinh (2(1-x)t)\sinh ht}  \label{ttb}
\end{eqnarray}
where $\chi _{\alpha }(h,t)=\sinh ((h-2x\rho _{\alpha }-2(1-x)\rho _{\alpha
}^{\vee })t);$ $x=b^{2}/(1+b^{2})$ and variable $w_{r}$ is defined by
Eq.(\ref{w}). The expression for vector $\Theta _{b}(C)$
can be easily derived
from Eqs.(\ref{DC}-\ref{th}). It can be written in terms of functions
$\chi_{\alpha }^{(0)}=
\chi _{\alpha }(0,t)$ and $\chi _{\alpha }^{(r)}=
\chi_{\alpha }(2h,t)$ in the form:

\begin{eqnarray}
\frac{\Theta _{b}(C)}{x(1-x)} &=&-\sum_{\alpha >0}\alpha
\int\limits_{0}^{\infty }\frac{dt\sinh ((1-x+\alpha ^{2}x/2)t)\chi _{\alpha
}(h,t)}{\sinh ((1-x)t)\,\cosh (\alpha ^{2}xt/2)\sinh ht}  \nonumber \\
&&+\sum_{k=0,r}\sum_{\alpha \in \mathbf{C}}4\alpha \int\limits_{0}^{\infty }
\frac{dt(\sinh ^{2}(2w_{k}t)-\sinh ^{2}((1-x)t))\chi _{\alpha }^{(k)}(t)}
{\sinh (\alpha ^{2}xt)\sinh (2(1-x)t)\sinh 2ht}  \label{tc}
\end{eqnarray}

The vector $\Theta _{b}$ can be expressed in terms of elementary 
functions (see
section 5) in two dual classical limits $b\rightarrow 0$ ($x=0$) and
$b\rightarrow \infty $ ($x=1$).
As it was explained in Ref.\cite{FO}, in these
two limits the values of the vector $\Theta _{b}$ determine the boundary 
values
$\phi (0)$ of the solutions $\phi (y)$ to classical boundary Toda equations
for the dual pair of ATTs. These solutions describe the classical vacuum
configurations i.e. correspond to the dual classical limits of the
correlation function:

\begin{equation}
\Phi _{b}(y)=\left\langle \widehat{\varphi }(x,y)\right\rangle
_{B}-\left\langle 0|\widehat{\varphi }|0\right\rangle   \label{FI}
\end{equation}
where the first term in the right hand side of this equations denotes the
VEV of the bulk field in boundary ATT and the second is the VEV of field
$\widehat{\varphi }$ in the theory defined on the whole plane (see 
section 2).

After rescaling and shifting (see Eq.(\ref{ACL})) the classical equations
which follow from action (\ref{atb}) have a form:
\begin{equation}
\partial _{y}^{2}\phi =m^{2}\sum_{i=0}^{r}n_{i}e_{i}\exp (e_{i}\cdot \phi );
\label{eqt}
\end{equation}
with the boundary condition at $y=0$:

\begin{equation}
\partial _{y}\phi =m\sum\limits_{i=0}^{r}\sqrt{n_{i}}A_{i}e_{i}\exp
(e_{i}\cdot \phi /2).  \label{yn}
\end{equation}
where the coefficients $A_{i}$ are defined by Eqs.(\ref{Bcl}-\ref{Cd}).

The limiting values of vector $\Theta _{b}$ determine the boundary
values $\phi (0)$ of the solution and Eq.(\ref{yn}) defines the
derivative $\partial_{y}\phi (0)$.  This gives us the possibility to
study Eq.(\ref{eqt}) numerically. The numerical analysis of this
equation shows that only for these boundary values smooth solutions
decreasing at infinity exist.

It is natural to expect that solution $\phi (y)$ to the Eqs.(\ref{eqt},\ref
{yn}) can be expressed in terms of tau-functions associated with
multi-soliton solutions to classical ATTs equations (see for example \cite
{BOWC}, \cite{FO} and references there). We postpone the explicit
construction of these solutions to section 5. Here we consider the classical
and quantum boundary ground state energies (BGSEs). The classical BGSE can
be expressed in terms of the solution $\phi (y)$ as follows:

\begin{eqnarray}
\mathcal{E}^{(cl)} &=&\frac{1}{4\pi b^{2}}[2m\sum_{i=0}^{r}\sqrt{n_{i}}
A_{i}\exp (e_{i}\cdot \phi (0)/2)  \nonumber \\
&&+\int\limits_{0}^{\infty }dy(\frac{1}{2}(\partial _{y}\phi
)^{2}+m^{2}\sum_{i=0}^{r}n_{i}(e^{e_{i}\cdot \phi }-1))].  \label{cle}
\end{eqnarray}
For some special values of parameters $w_{0,r}$ classical BGSEs are known.
If we put $w_{0}=w_{r}=1/2$ in Eqs.(\ref{Bdcl},\ref{Ccl}) for the boundary
conditions corresponding to $B_{r}^{\vee }$ and $C_{r}$ ATTs we obtain that
all coefficients $A_{i}$ vanish (Neumann boundary conditions). It means that
\ solution $\phi (y)$ also vanishes and:

\begin{equation}
\mathcal{E}^{(cl)}(B_{r}^{\vee })=\mathcal{E}^{(cl)}(C_{r})=0;\quad
w_{0}=w_{r}=1/2  \label{3c}
\end{equation}
Putting $w_{0}=w_{r}=0$ in boundary conditions (\ref{Bcl},\ref{Cd})
corresponding to $B_{r}$ and $C_{r}^{\vee }$ ATTs we find that all $A_{i}=
\sqrt{2/e_{i}^{2}}$ (classical version of dual to Neumann boundary
conditions). The explicit solutions $\phi (y)$ for these values of
coefficients $A_{i}$ were found for all ATTs in \cite{FO}. The corresponding
BGSEs are:

\begin{equation}
\mathcal{E}^{(cl)}(C_{r-1}^{\vee })=\mathcal{E}^{(cl)}(B_{r})=\frac{rm\cos
(\pi /4-\pi /4r)}{\pi b^{2}\cos (\pi /4r)};\quad w_{0}=w_{r}=0  \label{2c}
\end{equation}

To calculate $\mathcal{E}^{(cl)}$ for arbitrary values of boundary
parameters we can use the relation:

\begin{equation}
\partial _{w_{r}}\mathcal{E}^{(cl)}(G)=\frac{2m\sqrt{n_{i}}(\partial
_{w_{r}}A_{r}(G))}{4\pi b^{2}}\exp (e_{r}\cdot \phi (0)/2)  \label{pr}
\end{equation}
and similar equation with respect the parameter $w_{0}$ for
$G=C_{r},C_{r}^{\vee }$.
The boundary values $\phi (0)$ of the solution are
defined by the dual classical limits of vector $\Theta _{b}(G)$. All
boundary values of exponents:

\begin{equation}
E_{i}=\exp (e_{i}\cdot \phi (0)/2);\quad i=0,...,r  \label{ei}
\end{equation}
are given in section 6. The integration of the Eqs.(\ref{pr}) with ``initial
conditions`` (\ref{3c},\ref{2c}) will give us the classical BGSEs.

In the quantum case the BGSEs for $B_{r}$ and $C_{r}^{\vee }$ ATTs with dual
to Neumann boundary conditions (\ref{ND}) which in the quantum case are
characterized by the values of variables $w_{0}=w_{r}=x/2$ (and hence also
for $B_{r}^{\vee }$ and $C_{r}$ ATTs with Neumann boundary conditions:
$w_{0}=w_{r}=(1-x)/2$) were conjectured in \cite{FO}. Here we accept this
conjecture which is:

\begin{equation}
\mathcal{E}^{(q)}(B_{r})=\frac{m\cos (\pi /4-\pi /2h)}{4\sin (\pi x/2h)
\,\cos(\pi (1-x)/2h)};\quad h=2r-x.  \label{bb}
\end{equation}
To obtain the expression for $\mathcal{E}^{(q)}(B_{r}^{\vee })$ we should do
in Eq.(\ref{bb}) the substitution: $x\rightarrow 1-x$. The conjectured form
for $\mathcal{E}^{(q)}(C_{r}^{\vee })$ (with the same substitution for
$\mathcal{E}^{(q)}(C_{r})$) is also given by Eq.(\ref{bb}) where we should
take $h=2r+2-2x$.

To calculate BGSEs for arbitrary values of variables $w_{r},w_{0}$ we can
find from Eq.(\ref{atb}) that:

\begin{equation}
\partial _{w_{i}}\mathcal{E}^{(q)}(G)=(b+1/b)(\partial _{s_{i}}\nu _{i}(G))
\mathcal{G}_{G}(be_{i});\quad i=0,r  \label{qbe}
\end{equation}
The values of function $\mathcal{G}_{G}(a)$ at the special points $be_{i}$
can be calculated explicitly with a result:

\begin{equation}
\partial _{w_{r}}\mathcal{E}^{(q)}(B_{r})=-\frac{\pi m\sin (\pi w_{r}/h)}
{2\sqrt{2}h\sin (\pi x/2h)\sin (\pi (1-x)/h)};  \label{wrb}
\end{equation}

\begin{equation}
\partial _{w_{r}}\mathcal{E}^{(q)}(C_{r}^{\vee })=-\frac{\pi m\sin (\pi
w_{r}/h)\,\cos (\pi w_{0}/h)}{\sqrt{2}h\sin (\pi x/h)\,\sin (\pi (1-x)/h)};
\label{wr}
\end{equation}

\begin{equation}
\partial _{w_{0}}\mathcal{E}^{(q)}(C_{r}^{\vee })=-\frac{\pi m\sin (\pi
w_{0}/h)\,\cos (\pi w_{r}/h)}{\sqrt{2}h\sin (\pi x/h)\,\sin (\pi (1-x)/h)}.
\label{w0}
\end{equation}
The integration of these equations with ``initial conditions'' (\ref{bb})
gives:

\begin{equation}
\mathcal{E}^{(q)}(B_{r})=\frac{m\cos (\frac{\pi }{4}-\frac{\pi }{2h})}{4\sin
(\frac{\pi x}{2h})\cos (\frac{\pi (1-x)}{2h})}+
\frac{m(\cos (\frac{\pi w_{r}}{h})-
\cos (\frac{\pi x}{2h}))}{2\sqrt{2}\sin (\frac{\pi x}{2h})
\,\sin (\frac{\pi (1-x)}{h})};  \label{e}
\end{equation}

\begin{equation}
\mathcal{E}^{(q)}(C_{r}^{\vee })=\frac{m\cos (\frac{\pi }{4}-
\frac{\pi }{2h})}{4\sin (\frac{\pi x}{2h})
\,\cos (\frac{\pi (1-x)}{2h})}+\frac{m(\cos (\frac{\pi w_{r}}{h})
\,\cos (\frac{\pi w_{0}}{h})-
\cos ^{2}(\frac{\pi x}{2h}))}{\sqrt{2}
\sin (\frac{\pi x}{h})\,\sin (\frac{\pi (1-x)}{h})}  \label{2}
\end{equation}
The expressions for the BGSEs $\mathcal{E}^{(q)}(B_{r}^{\vee })$ and
$\mathcal{E}^{(q)}(C_{r})$ can be derived from these equations by the
substitution $x\rightarrow 1-x$. The classical BGSEs can be derived
from Eqs.(\ref{e},\ref{2}) as the main terms
($O(1/x)$ or $O(1/(1-x))$) of the asymptotics of the quantum values.

We note that $C_{1}^{\vee }$ and $C_{1}$ ATTs coincide with boundary
sinh-Gorgon model. In this case BGSE (\ref{2}) being written in terms of the
mass $m_{1}=\sqrt{8}m\sin (\pi /h)$ coincides with expression for boundary
ground state energy of this model proposed by Al.Zamolodchikov.
\cite{AlZ} (see also \cite{MRD}).

The nonperturbative check of conjectures (\ref{e},\ref{2}) can be made using
the boundary Thermodynamic Bethe Ansatz equations \cite{SSM}.
The kernels in these
nonlinear integral equations depend on the bulk S-matrices $S_{ij}(\theta)$
and the source (inhomogeneous) terms on the boundary $S$-matrices
$R_{j}(\theta )$. Both these $S$-matrices in ATTs are the pure phases.
The  boundary ground state energy can be expressed through
the main terms of the asymptotics of these phases at
$\theta\rightarrow \infty$ \cite{PDRT}.
In particular, boundary ground state energy in ATT can be expressed
in terms of the
mass $m_{j}$ of the particle $j$ multiplied by the ratio of Fourier
transforms $\Delta_j(\omega)$ and $\Delta_{jj}(\omega)$  of logarithms of
boundary $S$-matrix $R_{j}$ and bulk amplitude
$S_{jj}$ taken at $\omega =i$. To check our conjecture and to construct the
explicit solution of Eqs.(\ref{eqt},\ref{yn}) we need the information about
the boundary $S-$matrix.

\section{Boundary S-matrix}

The boundary $S-$matrix (reflection coefficient) for the particle $j$ with
mass $m_{j}$ (\ref{spr}) in integrable boundary ATTs can be defined as:

\begin{equation}
|j,-\theta \rangle _{B,out}=R_{j}(\theta )|j,\theta \rangle _{B,in}
\label{BS}
\end{equation}
where $\theta $ is a rapidity of particle $j$.

To be consistent with a bulk $S-$ matrix $S_{ij}(\theta )$ the reflection
coefficients $R_{j}(\theta )$\ should satisfy the boundary bootstrap
equations \cite{FK},\cite{GZ}. These equations can be written in terms of
the fusion angles $\theta _{ij}^{k}$ which determine the position of the
pole corresponding to the particle $k$ in the amplitude $S_{ij}(\theta )$.
We denote as $\overline{\theta }_{ij}^{k}=\pi -\theta _{ij}^{k}$, then the
consistency equations can be written as:

\begin{equation}
R_{k}(\theta )=R_{i}(\theta -i\overline{\theta }_{ik}^{j})R_{j}(\theta +
i\overline{\theta }_{jk}^{i})S_{ij}(2\theta +i\overline{\theta }_{jk}^{i}-
i\overline{\theta }_{ik}^{j}).  \label{Bb}
\end{equation}
The ``crossing unitarity'' relation \cite{GZ} impose additional condition
for boundary $S$-matrix:

\begin{equation}
R_{j}(\theta )R_{j}(\theta +i\pi )=S_{jj}(2\theta )  \label{Cu}
\end{equation}

The reflection coefficient in ATT is a pure phase $R_{j}=\exp (i\delta
_{j}(\theta ))$.

We give here the conjecture for minimal solutions to these equations which
are consistent with boundary ground state energies (\ref{e}.\ref{2}). We
represent the boundary amplitudes $R_{j}(\theta )$ in the form of Fourier
integrals which is convenient for TBA analysis.

For the particles $j$ with the masses $m_{j}=\sqrt{8}m\sin (\pi j/h)$ (see
Eq.(\ref{spr})) the reflection coefficients can be written as:

\begin{equation}
R_{j}(\theta )=\exp \left( i\int\limits_{0}^{\infty }\frac{dt}{t}\sin \left(
2h\theta t/\pi \right) \left[ \Delta _{j}(G,t)+2\right] \right)  \label{del}
\end{equation}
where

\begin{eqnarray}
\Delta _{j}(G,t) &=&-\frac{4\sinh (1-x)t\,\sinh (h+x)t\,\sin jt\,\cos 
(h/2-j)t}
{\sinh t\,\cosh ht/2\,\cosh ht}+  \nonumber \\
&&+\frac{8\sinh (1-x)t\,\cosh xt\,\sinh jt\,\sinh (j-1)t}{\sinh 
2t\,\cosh ht}+
\psi_{j}(G,t).  \label{de}
\end{eqnarray}
For the dual pair of $\ B_{r}$ and $B_{r}^{\vee }$ of integrable boundary
ATTs the functions $\psi _{j}$ ($j=1,...,r-1$) depend on one parameter
$w_{r} $ and have a form:

\begin{equation}
\psi _{j}(B_{r},t)=\frac{8\sinh 2jt\,\cosh xt\,\sinh (w_{r}+x/2)t\,\sinh
(w_{r}-x/2)t}{\sinh 2t\,\cosh ht}.  \label{deb}
\end{equation}
For the particle $r$ with mass $\sqrt{2}m$ the solution is:

\begin{eqnarray}
\Delta _{r}(B_{r},t) &=&-\frac{2\sinh (1-x)t\,\sinh 2rt\,\sin (h/2+1)t}
{\sinh2t\,\cosh ht/2\,\cosh ht}  \label{dnb} \\
&&+\frac{4\sinh 2rt\,\sinh (w+x/2)t\,\sinh (w-x/2)t}{\sinh 2t\,\cosh ht}.
\nonumber
\end{eqnarray}
The dual pair of $C_{r}^{\vee }$ and $C_{r}$ boundary ATTs is characterized
by the functions $\psi _{j}(C_{r}^{\vee },t)$ ($j=1,...,r$) which depend on
two parameters $w_{0},w_{r}$ as follows:

\begin{equation}
\psi _{j}(C_{r}^{\vee },t)=\frac{4\sinh 2jt(\cosh (2w_{1}t)\,\cosh
(2w_{2}t)-\cos ^{2}(xt))}{\sinh 2t\,\cosh ht}.  \label{dec}
\end{equation}
All amplitudes $R_{j}$ in this theory can be obtained using Eq.(\ref{Bb})
from the amplitude $R_{1}$ corresponding to the lightest particle $m_{1}$.
It follows from Eqs.(\ref{del},\ref{de},\ref{dec}) that amplitude
$R_{1}(C_{r}^{\vee })$ can be written in terms of functions $(z)=
\frac{\sinh(\theta /2+i\pi z/2h)}{\sinh (\theta /2-i\pi z/2h)}$ as:

\begin{equation}
R_{1}(\theta )=\frac{(x-1)(h/2+x)(h-x)(h/2+1-x)(h/2)(h/2-1)}
{(h-1)(h/2+w_{+})(h/2-w_{+})(h/2-w_{-})(h/2+w_{-})}  \label{r1c}
\end{equation}
here $w_{\pm }=w_{r}\pm w_{0}$. The amplitude $R_{1}(B_{r})$ can be obtained
from this equation if we put $w_{0}=x/2$.

We give here also the conjecture for boundary $S$-matrix in $BC_{r}$ ATT
which flows from boundary conditions (\ref{BC1}) to boundary conditions
(\ref{BC2}). The boundary parameters $w_{0}$ and $w_{r}$ in these equations
correspond to dual classical limits of the same parameter which we
denote as
$w$. Then amplitudes $R_{j}(\theta )$ ($j=1,..,r$) are characterized by
functions $\psi _{j}(BC_{r},t)$ which are:

\begin{equation}
\psi _{j}(BC_{r},t)=\frac{4\sinh 2jt(\,\cosh t\,\cosh (2wt)-\cos ^{2}(xt))}
{\sinh 2t\,\cosh ht}  \label{debc}
\end{equation}
This conjecture is consistent with the following form of boundary ground
state energy

\begin{equation}
\mathcal{E}^{(q)}(BC_{r})=\frac{m\cos (\frac{\pi }{4}-\frac{\pi }{2h})}
{4\sin (\frac{\pi x}{2h})\,\cos (\frac{\pi (1-x)}{2h})}+
\frac{m(\cos (\frac{\pi w}{h})\,\cos (\frac{\pi }{2h})-
\cos ^{2}(\frac{\pi x}{2h}))}
{\sqrt{2}\sin (\frac{\pi x}{h})\,\sin (\frac{\pi (1-x)}{h})}.  \label{v}
\end{equation}

The boundary $S-$matrices described above correspond to the minimal
solutions of Eqs.(\ref{Bb},\ref{Cu}) which are consistent with GSESs that
were derived in the previous section. For $\left| w_{_{\pm }}\right| <1$ and
for $|w_{r}|<1/2$ the amplitudes $R_{j}(\theta )$ have no poles
corresponding to the bound states of the particles $j$ with boundary. The
only poles with positive residues which appear in these functions are the
poles at $\theta =i\pi /2$

\begin{equation}
R_{j}(\theta )=\frac{iD_{j}^{2}(G)}{\theta -i\pi /2}  \label{dsq}
\end{equation}
corresponding to the particle boundary coupling with zero binding energy.
For example, the lightest particle whose amplitude possesses this pole is
the particle $m_{1}$. It is easy to derive from Eq.(\ref{r1c}) that
corresponding residue has a form:

\begin{equation}
D_{1}^{2}(C_{r}^{\vee })=\frac{2\cot \left( \frac{\pi x}{2h}\right) \cot
\left( \frac{\pi (1-x)}{2h}\right) \tan ^{2}\left( \frac{\pi w_{+}}{2h}
\right) \tan ^{2}\left( \frac{\pi w_{-}}{2h}\right) }{\tan \left( \frac{\pi
}{2h}\right) \cot \left( \frac{\pi }{4}-\frac{\pi }{2h}\right) \cot \left(
\frac{\pi }{4}-\frac{\pi (1-x)}{2h}\right) \tan \left( \frac{\pi }{4}-
\frac{\pi x}{2h}\right) }.  \label{r}
\end{equation}
The residue $D_{1}^{2}(B_{r})$ can be obtained from this equation if we put
$w_{0}=x/2$.

It was shown in Ref.\cite{FO} that the classical limits of the residues
$D_{j}$:

\begin{equation}
d_{j}(G)=\lim_{x\rightarrow 0}\sqrt{\pi x(1-x)}D_{j}(G);\quad d_{j}(G^{\vee
})=\lim_{x\rightarrow 1}\sqrt{\pi x(1-x)}D_{j}(G).  \label{dgg}
\end{equation}
play the crucial role for the construction of the explicit solutions
to Eqs.(\ref{eqt}\ref{yn}).
Namely, they determine ``one particle'' contributions to
the boundary solution $\phi (y)$ which describes the classical vacuum
configuration:

\begin{equation}
\phi (y)=\sum_{j=1}^{r}d_{j}\xi _{j}\exp (-m_{j}y)+...  \label{as}
\end{equation}
here $\xi _{j}$ are the eigenvectors of mass matrix (\ref{mab}) ($m^{2}M\xi
_{j}=m_{j}^{2}\xi _{j}$), satisfying the conditions:

\begin{equation}
\xi _{i}\cdot \xi _{j}=\delta _{ij};\quad \xi _{i}\cdot \rho ^{\vee }\geq 0.
\label{mv}
\end{equation}
For example, the coefficient $d_{1}$ corresponding to the lightest particle
$m_{1}$ in expansion (\ref{as}) and defining the main term of the
asymptotics
at $y\rightarrow \infty $ can be extracted from Eq.(\ref{r}) and has a form:

\begin{equation}
d_{1}(C_{r}^{\vee })=2\sqrt{H}\tan \left( \frac{\pi }{4}-\frac{\pi }{2H}
\right) \cot \left( \frac{\pi }{2H}\right) \tan \left( \frac{\pi w_{+}}{2H}
\right) \tan \left( \frac{\pi w_{-}}{2H}\right)  \label{11}
\end{equation}

\begin{equation}
d_{1}(C_{r})=2\sqrt{H}\cot \left( \frac{\pi }{2H}\right)
\tan \left( \frac{\pi w_{+}}{2H}\right)
\tan \left( \frac{\pi w_{-}}{2H}\right)  \label{12}
\end{equation}
where $H(C_{r}^{\vee })=2r+2$ and $H(C_{r})=2r$. The coefficient
$d_{1}(B_{r})$ ($d_{1}(B_{r}^{\vee })$) can be obtained from Eq.(\ref{11})
(Eq.(\ref{12})) if we put there $w_{0}=0$ ($w_{0}=1/2$).

The coefficients $d_{j}$ as well as the boundary values $\phi (0)$ fix
completely the solution to the Eqs.(\ref{eqt},\ref{yn}). They determine the
contribution of the zero modes of the linearized Eq.(\ref{eqt}) and make it
possible to develop in a standard way the regular expansion of the solution
$\phi (y)$ at large distances. If our scattering theory is consistent with
conformal perturbation theory this expansion should converge to the boundary
value $\phi (0)$. In the next section we use the coefficients $d_{i}$ to
construct the exact solution and to check this consistency condition.

\section{Boundary Solutions}

It is natural to assume that boundary vacuum solutions can be expressed in
terms of tau-functions associated with multisoliton solutions of classical
ATTs equations. For all cases that we consider below it is convenient to
represent these solutions in the form:

\begin{equation}
\phi (y)=-\frac{1}{2}\sum_{k=0}^{r}n_{k}e_{k}\log \tau _{k}(y);\quad \tau
_{k}(y)\rightarrow 0,~y\rightarrow \infty  \label{t}
\end{equation}
where numbers $n_{k}$ and roots $e_{k}$ characterize the corresponding ATT.
If we chose the standard basis of roots:

\begin{equation}
e_{k}\cdot \phi =\phi _{k}-\phi _{k+1};\quad k=1,...,r-1  \label{SB}
\end{equation}
and take the roots $e_{0}$ and $e_{r}$ in accordance with extended (affine)
Dynkin diagram, we obtain that:

\begin{equation}
\phi _{k}=\log (\tau _{k-1}/\tau _{k});\quad k=1,...,r  \label{fii}
\end{equation}
The classical boundary ground state energy can be expressed in terms of
boundary values of $\tau $-functions \cite{BOWC} as:

\begin{equation}
\mathcal{E}^{(cl)}(G)=\frac{H}{4\pi b^{2}}\left( A_{k}\sqrt{n_{k}}\exp
(e_{k}\phi (0)/2)+\frac{\tau _{k}^{\prime }(0)}{\tau _{k}(0)}\right)
;~k=0,...,r.  \label{en}
\end{equation}

All functions $\tau _{k}$ corresponding to multisoliton solutions of ATT
equations are given by finite order polynomials in the variable

\begin{equation}
Z_{j}(y)=\exp (-m_{j}y).  \label{Z}
\end{equation}
The coefficients of these polynomials are completely fixed by the equations
of motion and by the asymptotics (\ref{as}). The construction of these
polynomials for boundary solutions in simply laced ATTs was described in
details in Ref.\cite{FO}. In all cases described below it is very similar to
construction for boundary solution in $D_{r}$ ATT. In particular, to
simplify the form of tau-functions it is convenient similar to $D_{r}$ case
to introduce the parameters $t_{j}$ which for $j\leq r$ are related with
coefficients $d_{j}$ as:

\begin{equation}
t_{j}=\frac{d_{j}}{2\sqrt{H}\sin (\pi j/H)}  \label{tj}
\end{equation}
This normalization is useful, because
the following relations between the roots $e_{k}$ (that appear
in Eq.(\ref{t}) for $\phi(y)$) and eigenvectors of mass
matrix $\xi _{j}$ (that describe
 ``one particle'' contributions (\ref{as}) to the solution) are valid:

\begin{equation}
-\sum_{k=0}^{r}n_{k}e_{k}\cos (\pi j(2k+\kappa (G))/H)=2\sqrt{H}\sin (\pi
j/H)\xi _{j}  \label{eks}
\end{equation}
where $\kappa (B_{r})=\kappa (B_{r}^{\vee })=-1;$ $\kappa
(C_{r}^{\vee })=1$ and $\kappa (C_{r})=0$.

Below we give the boundary values of the solutions that follow from
classical limits of vector $\Theta _{b}(G)$, the explicit expressions for
coefficients $t_{j}$ that follow from the boundary $S-$matrices and
construct the exact boundary vacuum solutions.

\subsection{B$_{r}$ boundary solution}

To specify the boundary values of the solutions it is convenient to
introduce the functions:

\begin{equation}
q_{l}^{2}(w)=\frac{(\cos (\frac{2\pi w}{H})+\cos (\frac{2\pi l}{H}))}
{(1+\cos (\frac{2\pi l}{H}))}
\prod_{i=0}^{l-1}\frac{(\cos (\frac{2\pi w}{H})+
\cos (\frac{2\pi (l-2i)}{H}))}{(\cos (\frac{2\pi w}{H})+\cos (\frac{2\pi
(l-2i-1)}{H}))}  \label{h}
\end{equation}
and numbers:

\begin{equation}
p_{l}^{2}=\prod_{i=1}^{l}\frac{\cos (\pi (l+1-2i)/H)}{\cos (\pi (l-2i)/H)}.
\label{p}
\end{equation}
Then boundary values $E_{i}$ defined by Eq.(\ref{ei}) for $B_{r}$ solution
can be derived from vector $\Theta _{0}(B)$ (see Eq.(\ref{ttb})) and\
written as follows:

\begin{eqnarray}
E_{0} &=&E_{1}=\frac{\sqrt{8}\cos (\pi w/H)}{H\sin (\pi /H)};~E_{r}=
\frac{\sqrt{2}\sin (\pi w/H)}{\sin (\pi w/2)\,\sin (\pi /H)};  \nonumber \\
E_{k} &=&\frac{\cos (\pi k/H)\cos (\pi
(r+1-k)/H)p_{k}^{2}p_{r+1-k}^{2}q_{k-1}(w_{r})}{\cos (\pi (2k-1)/2H)\,\cos
(\pi (2r-2k+1)/2H)q_{k-1}(0)}  \label{eb}
\end{eqnarray}
where $k=2,..,r-1$ and $H=2r$.

The analysis of the boundary $S-$matrix (\ref{del}-\ref{dnb}) gives that
for
$r>1$ the amplitudes $R_{j}(\theta )$ with $j=1,...r-1$ have a pole at
$\theta =i\pi /2$. The amplitude $R_{r}(\theta )$ has no such pole and
particle $m_{r}$ does not contribute to the solution. This selection rule
follows from $\mathbf{Z}_{2}$ symmetry of affine Dynkin diagram and
integrable boundary conditions (\ref{Bcl}). The parameters $t_{j}$
(\ref{tj})
can be derived from explicit form of amplitudes $R_{j}$ (see section 4)
and written as:

\begin{equation}
t_{j}=\frac{\tan (\frac{\pi }{4}-\frac{\pi j}{2H})}{2\cos ^{2}(\frac{\pi j}
{2H})}\prod_{i=0}^{j-1}\frac{\tan ^{2}(\frac{\pi (j-1-2i+w_{r})}{2H})}
{\tan^{2}(\frac{\pi (i+1)}{2H})}.  \label{tjb}
\end{equation}
Parameters $t_{j}$ and functions $Z_{j}=\exp (-ym\sqrt{8}\sin (\pi j/H))$
are defined for $j\leq r-1$. To write the solution in the most short form it
is convenient, however, to continue these values to $j\leq H-1$. To continue
parameters $t_{j}$ we can use Eq.(\ref{tjb}). In this way we obtain:

\begin{equation}
t_{H-j}=-t_{j};\quad Z_{H-j}(y)=Z_{j}(y).  \label{tZ}
\end{equation}

The exact vacuum solution to Eqs.(\ref{eqt},\ref{yn}) in $B_{r}$ ATT with
boundary conditions (\ref{Bcl})\ can be written in the form (\ref{t})
where:
\begin{equation}
\tau _{k}(y)=\sum_{\sigma _{1}=0}^{1}...\sum_{\sigma
_{H-1}=0}^{1}\prod_{j=1}^{H-1}\Omega _{k}^{j\sigma _{j}}(t_{j}Z_{j})^{\sigma
_{j}}\prod_{m<n}^{H-1}\left( \frac{\sin \left( \frac{\pi (m-n)}{2H}\right) }
{\sin \left( \frac{\pi (m+n)}{2H}\right) }\right) ^{2\sigma _{m}\sigma _{n}}
\label{sol}
\end{equation}
where for $B_{r}$ solution $\Omega _{k}=\exp (i\pi (2k-1)/H)=\exp (i\pi
(2k-1)/2r)$.

It is easy to derive from Eqs.(\ref{tZ},\ref{sol}) and (\ref{tj},\ref{eks})
that ``one particle'' contributions to the solution $\phi (y)$ are given by
Eq.(\ref{as}).
It can be also checked that boundary values of
the solution $\phi (y)$ defined by
Eqs.(\ref{t},\ref{sol}) coincide with vector $\Theta _{0}(B)$ defined by
Eq.(\ref{eb}) and the classical boundary ground state energy calculated
using
Eq.(\ref{en}) coincides with the main term of the asymptotics ($O(1/x)$) of
Eq.(\ref{e}).

\subsection{B$_{r}^{\vee }$ boundary solution}

The boundary values for this solution are determined by vector
$\Theta_{\infty }(B)$ (see Eq.(\ref{ttb})) and are:

\begin{equation}
E_{k}^{2}=\frac{\cos (2\pi w_{r}/H)+\cos (\pi (2j-1)/H)}{\cos (\pi /H)+\cos
(\pi (2j-1)/H)};~E_{r}=\frac{\sin (\pi w_{r}/H)}{\sin (\pi w_{r})
\,\sin (\pi/2H)}.  \label{bde}
\end{equation}
where $k=0,...,r-1$ and $H=2r-1$.

The residues in the poles of the amplitudes $R_{j}(\theta )$ ($j=1,...,r-1$)
at $\theta =i\pi /2$ determine the following expression for parameters
$t_{j} $:

\begin{equation}
t_{j}=\frac{1}{2\cos ^{2}(\frac{\pi j}{2H})}\prod_{i=1}^{j}
\frac{\tan (\frac{\pi (2w_{r}+2i-1)}{4H})
\tan (\frac{\pi (2w_{r}-2i+1)}{4H})}
{\tan ^{2}\left(
\frac{\pi i}{2H}\right) }.  \label{tbdu}
\end{equation}
Parameters $t_{j}$ and functions $Z_{j}=\exp (-ym\sqrt{8}\sin (\pi j/H))$
are defined for $j\leq r-1$. To write the solution in the form (\ref{sol})
we should continue these values to $j\leq H-1$. To continue parameters
$t_{j} $ we can use Eq.(\ref{tbdu}). In this way we obtain:
$t_{H-j}=-t_{j}$; $Z_{H-j}(y)=Z_{j}(y)$.

The exact vacuum solution to Eqs.(\ref{eqt},\ref{yn}) in $B_{r}^{\vee }$ ATT
with boundary conditions (\ref{Bdcl})\ can be written in the form (\ref{t})
where functions $\tau _{k}(y)$ have a form (\ref{sol}) with $t_{j}$ defined
by Eq.(\ref{tbdu}) and $\Omega _{k}=\exp (i\pi (2k-1)/H)=\exp (i\pi
(2k-1)/(2r-1))$.

It can be checked that boundary values of this solution coincide with
vector
$\Theta _{\infty }(B)$ defined by Eq.(\ref{bde}) and the classical boundary
ground state energy calculated using Eq.(\ref{en}) coincides with the main
term of the asymptotics ($O(1/(1-x))$) of \ Eq.(\ref{e}).

\subsection{C$_{r}^{\vee }$ boundary solution}

The boundary values of this solution are determined by vector $\Theta
_{\infty }(C)$ (see Eq.(\ref{tc})) and have a form:

\begin{eqnarray}
E_{0} &=&\frac{\sqrt{2}\sin (\pi w_{0}/H)\,\cos (\pi w_{r}/H)}
{\sin (\pi w_{0}/2)\,\sin (\pi /H)};~E_{r}=
\frac{\sqrt{2}\sin (\pi w_{r}/H)\,\cos (\pi w_{0}/H)}
{\sin (\pi w_{r}/2)\,\sin (\pi /H)};  \nonumber \\
E_{k} &=&\frac{\cos (\pi (k+1)/H)\,\cos (\pi
(r-k+1)/H)p_{k+1}^{2}p_{r+1-k}^{2}q_{k}(w_{1})q_{r-k}(w_{2})}
{\cos (\pi(2k+1)/2H)
\,\cos (\pi (2r-2k+1)/2H)q_{k}(0)q_{r-k}(0)}  \label{3}
\end{eqnarray}
where $k=1,...,r-1$ and $H=2r+2$.

The analysis of boundary $S-$ matrix (\ref{de},\ref{dec}) gives that for
$w_{\pm }\neq 0$ ($w_{\pm }$ $=w_{r}\pm w_{0}$)
all amplitudes $R_{j}(\theta) $
have a pole at $\theta =i\pi /2$ and all $r$ particles with masses
$m_{j}=\sqrt{8}\sin (\pi j/H)$ contribute to the solution. The parameters
$t_{j}$ can be extracted from the explicit form of
these amplitudes and have a form:

\begin{equation}
t_{j}=\frac{\tan (\frac{\pi }{4}-\frac{\pi j}{2H})}
{2\cos ^{2}(\frac{\pi j}{2H})}\prod_{i=0}^{j-1}
\frac{\tan (\frac{\pi (j-1-2i+w_{+})}{2H})
\tan (\frac{\pi (j-1-2i+w_{-})}{2H})}
{\tan ^{2}(\frac{\pi (i+1)}{2H})}.  \label{tjc}
\end{equation}
Parameters $t_{j}$ and functions $Z_{j}=\exp (-ym\sqrt{8}\sin (\pi j/H))$
are defined for $j\leq r$. To write the solution in the form (\ref{sol}) it
is necessary to continue these values to $j\leq H-1$. To continue
parameters
$t_{j}$ we can use Eq.(\ref{tbdu}). In this way we \ obtain:
$t_{H-j}=-t_{j}; $ $Z_{H-j}(y)=Z_{j}(y)$.

The vacuum solution in $C_{r}^{\vee }$ ATT with boundary conditions
(\ref{Cd}) can be written in the form (\ref{t})
(with all $n_{k}=2$) where functions
$\tau _{k}(y)$ have a form (\ref{sol}) with $t_{j}$ defined by
Eq.(\ref{tjc}) and
$\Omega _{k}=\exp (i\pi (2k+1)/H)=\exp (i\pi (2k+1)/(2r+2))$.

It can be checked that boundary values of this solution coincide with
vector
$\Theta _{\infty }(C)$ defined by Eq.(\ref{3}) and the classical boundary
ground state energy calculated using Eq.(\ref{en}) coincides with the main
term of the asymptotics ($O(1/x)$) of \ Eq.(\ref{2}).

\subsection{C$_{r}$ boundary solution}

The boundary values of this solution are defined by vector $\Theta _{0}(C)$
and have a form:

\begin{eqnarray}
E_{0} &=&\frac{2\sin (\pi w_{0}/H)\cos (\pi w_{r}/H)}{\sin (\pi w_{0})\,\sin
(\pi /H)};~E_{r}=\frac{2\sin (\pi w_{r}/H)\,\cos (\pi w_{0}/H)}{\sin (\pi
w_{r})\,\sin (\pi /H)};  \nonumber \\
E_{k}^{2} &=&\frac{(\cos (2\pi w_{r}/H)+\cos (2\pi k/H))(\cos (2\pi
w_{0}/H)-\cos (2\pi k/H))}{(\cos (\pi /H)+\cos (2\pi k/H))(\cos (\pi
/H)-\cos (2\pi k/H))}  \label{ec}
\end{eqnarray}
where $k=1,...,r-1$ and $H=2r$.

The parameters $t_{j}$ that follow from the boundary $S-$matrix are:

\begin{equation}
t_{j}=\frac{1}{2\cos ^{2}(\frac{\pi j}{2H})}\prod_{i=0}^{j-1}
\frac{\tan (\frac{\pi (j-1-2i+w_{+})}{2H})
\tan (\frac{\pi (j-1-2i+w_{-})}{2H})}{\tan
^{2}(\frac{\pi (i+1)}{2H})}.  \label{tjcd}
\end{equation}
Parameters $t_{j}$ and functions $Z_{j}=\exp (-ym\sqrt{8}\sin (\pi j/H))$
are defined for $j\leq r$. To write the solution in the form (\ref{sol}) it
is necessary to continue these values to $j\leq H-1$. To continue
parameters
$t_{j}$ we can use Eq.(\ref{tjcd}). In this way we \ obtain:
$t_{H-j}=t_{j};$
$Z_{H-j}(y)=Z_{j}(y)$.

The vacuum solution in $C_{r}$ ATT with boundary conditions (\ref{Ccl})\ can
be written in the form (\ref{t}) where functions $\tau _{k}(y)$ have a form
(\ref{sol}) with $t_{j}$ defined by Eq.(\ref{tjcd}) and $\Omega _{k}=\exp
(i2\pi k/H)=\exp (i\pi k/r)$.

It can be checked that boundary values of this solution coincide with
vector
$\Theta _{0}(C)$ defined by Eq.(\ref{ec}) and the classical boundary ground
state energy calculated using Eq.(\ref{en}) coincides with the main term of
the asymptotics ($O(1/(1-x))$) of \ Eq.(\ref{2}).

\subsection{BC$_{r}$ boundary solution}

In classical case we have two types of integrable boundary conditions (\ref
{BC1}) and (\ref{BC2}). The solution in both cases can be derived in a
standard way from the scattering data described in the previous section.
However, we can obtain these both solutions by the reduction of
$C_{2r}^{\vee }$ and $B_{r+1}^{\vee }$ solutions with respect to the
symmetries of corresponding affine Dynkin diagrams (see Ref.\cite{FO}) and
integrable boundary conditions.

To derive the solution corresponding to boundary conditions (\ref{BC1}) we
should take the $C_{2r}^{\vee }$ solution and put there $w_{0}=w_{r}$. Then
$\mathbf{Z}_{2}$ symmetry of affine Dynkin diagram and boundary conditions
manifest itself in vanishing all coefficients $t_{j}$ with odd $j_{.}.$ The
tau-functions in this case possess the symmetry $\tau _{j}=\tau _{2r-j}$. In
the standard basis of roots (\ref{SB}) and $e_{0}\cdot \phi =-\phi _{1}$,
$e_{r}\cdot \phi =2\phi _{r}$ the solution can be written in the form
(\ref{fii}) where $\tau _{0},...,\tau _{r}$ are the tau-functions for
$C_{2r}^{\vee }$ solution.
The corresponding classical boundary ground state
energy $\mathcal{E}^{(cl)}(BC_{r})=\mathcal{E}^{(cl)}(C_{2r}^{\vee })/2$.

To derive the solution corresponding to boundary conditions (\ref{BC2}) we
can fold the symmetry $\tau _{0}=\tau _{1}$ of $B_{r+1}^{\vee }$ solution.
In this way we obtain that $BC_{r}$ solution can be written in the form
(\ref
{fii}) where $\tau _{k}(BC_{r})=\tau _{k+1}(B_{r+1}^{\vee }).$ The
corresponding classical boundary ground state energy coincides with that
for
$B_{r+1}^{\vee }$ solution.

\section{Concluding Remarks}

In the previous sections we calculated one-point functions and conjectured
boundary scattering theories and boundary ground state energies for the
parametric families of integrable boundary Toda theories. The exact
solutions corresponding to dual semiclassical limits of correlation function
(\ref{FI}) were derived. These solutions were constructed using only
boundary scattering data. The boundary values of these solutions are in
exact agreement with the same values derived from one-point functions. This
gives us the test for the consistency between $S-$matrix and CFT data.

\textit{1.\/} To make the similar test for arbitrary values of the coupling
constant we can express the long distance expansion for correlation
function
$\Phi _{b}(y)$ in terms of the boundary $S-$matrices $R_{j}$ and form
factors of ATTs (see Ref.\cite{DPW} for details). This expansion should
converge to the boundary values which can be extracted from one-point
functions, boundary conditions and equations of motion. Namely, $\Phi
_{b}(0)=\Theta _{b}$ and

\[
\partial _{y}\Phi _{b}|_{y=0}=2\pi x\sum_{i=0}^{r}\nu _{i}e_{i}\mathcal{G}
(be_{i});~\partial _{y}^{2}\Phi _{b}|_{y=0}=
4\pi x\sum_{i=0}^{r}\mu_{i}e_{i}
\mathcal{G}(2be_{i}).
\]
We suppose to do this test in the subsequent publications.

\textit{2.\/} Boundary solutions constructed above correspond to the case when
all signs $\varepsilon _{i}$ in Eq.(\ref{sig}) are positive. Together with
the results of Ref.\cite{FO} they give rather complete description of
classical vacuum configurations for this case in integrable boundary ATTs.
The situation with different signs $\varepsilon _{i}$ is more sophisticated.
For some choice of these signs the solution develops the singularity at the
boundary but gives the finite boundary ground state energy \cite{BOWC},\cite
{BPR}. Sometimes the minimum of this energy can not be achieved at the
static boundary solution. It looks interesting to analyze carefully how
these phenomena can be consistent with classical and quantum 
integrability. We
think that this problem needs the further study.

\textit{3.\/} In this paper we did not consider the boundary excited states. These
states appear for $|w_{\pm }|>1$ in $C_{r},C_{r}^{\vee }$ and for
$|w_{\pm}|>1/2$ in $B_{r},B_{r}^{\vee }$ boundary ATTs.
They manifest themselves in
the poles of boundary amplitudes $R_{j}(\theta )$ and can be considered as
the bound states of particles with a boundary. The boundary $S$-matrices
for this states and their spectrum can be derived by boundary bootstrap
method. In the classical limit these states can be seen as the boundary
breather solutions. The quantization of these solutions \cite{CT},\cite{CDE}
should be consistent with spectrum of boundary states. This gives an
additional test for boundary $S$-matrix. Another interesting problem related
with excited states is the calculation of the expectation values of the
boundary fields at these states. In the classical limit these expectation
values can be found from the boundary breather solutions. The problem of the
quantization of these expectation values as well as other problems mentioned
in this section we suppose to discuss in subsequent publications.

\bigskip
\begin{center}
\textbf{Acknowledgment}
\end{center}
We are grateful to Al. Zamolodchikov for useful
discussions.  E.O. warmly thanks A. Neveu, director of L.P.M.,
University of Montpellier II, and all his colleagues for the kind
hospitality at the Laboratory.  This work supported by part by the EU
under contract ERBFRMX CT 960012 and grant INTAS-OPEN-00-00055

\end{document}